\documentclass{article}
\usepackage{epsfig}

\begin{document}
\title{ A fitting function for the SNAP data }
\author{O.~V.~Lutchenko\footnote{olutchen@ip.rsu.ru}, G.~M.~Vereshkov}
\date{ Research Institute of Physics, \protect\\
Rostov State University, Stachki 194, Rostov-on-Don, 344090, Russia}
\maketitle

\begin{abstract}
A mathematically motivated fitting function for analyzing
supernovae distance-redshift data is suggested. It is
pointed out that this function can be useful in analyzing SNAP
data. To illustrate our method, we analyze current
experimental data which includes 78 supernovae and 20 radio galaxies.
From these data we obtain vacuum equation of state in a model
independent way. We show that the flat Universe model with a
cosmological constant is in a good agreement with observations.
\end{abstract}

\section{Introduction}
There is now strong evidence from observations \cite{r98}, \cite{p99},
\cite{d02} that the Universe is dominated by a nearly homogeneous component
--- the dark energy --- which is causing the cosmic expansion to accelerate
\cite{peebles02}, \cite{turner02}. At present, the distance
(magnitude)-redshift relation data is the main source of
information about dark energy. The new improved data about dark
energy will be obtained in the nearest future from SNAP experiment.
The SNAP satellite
will observe roughly 2000 supernovae (SNe) a year for three years with
very precise magnitude measurements out to a redshift of $z=1.7$
\cite{aldering02}. The problem is to establish the energy and
pressure as a function of time. In this paper we will focus on
the new method for analyzing experimental data which can be useful
in solving the problem.

The formal solution of this problem is well known. First of all, one needs
to obtain the distance-redshift function $r(z)$ from experimental
data. This function is usually referred to as a dimensionless coordinate
distance function. This procedure was carried out in \cite{r98}, \cite{p99} for
SNe and in \cite{daly} for radio galaxies (RGs). Then it is
necessary to find the first two derivatives of this function:
\begin{equation}
\begin{array}{c}
\displaystyle
r_{exp}(z)=\int_0^z\frac{dz}{\sqrt{\Omega_M(1+z)^3-(\Omega-1)(1+z)^2+\tilde
\varepsilon_{vac}(z)}},
\\[6mm]
\displaystyle r'_{exp}(z)\equiv \frac{dr_{exp}(z)}{dz},\qquad r''_{exp}(z) \equiv
\frac{d^2r_{exp}(z)}{dz^2}.
\end{array}
\label{in}
\end{equation}
Now, dark energy (vacuum) equation of state can be expressed through
experimental data:
\begin{equation}
\begin{array}{l}
\displaystyle
\frac{\varepsilon_{vac}(a)}{\rho_c}\equiv \tilde \varepsilon_{vac}(a)=
\left[ \frac{1}{\left(r'_{exp}(z)\right)^{2}}+(\Omega -1)(1+z)^2-\Omega_M(1+z)^3\right]_{z=1/a-1},
\\[7mm]
\displaystyle \frac{p_{vac}(a)}{\rho_c}\equiv \tilde p_{vac}(a)=
-\left[ \frac{1}{\left(r'_{exp}(z)\right)^{2}}+
\frac{2(1+z)r''_{exp}(z)}{3\left(r'_{exp}(z)\right)^{3}}
+\frac{(\Omega -1)(1+z)^2}{3}\right]_{z=1/a-1},
\end{array}
\label{ep1}
\end{equation}
The total energy density parameter $\Omega$ and the total matter density
$\Omega_M$ are obtained from independent observations.
Recent CMB measurements by WMAP give $\Omega = 1.02\pm 0.02,\;\Omega_M=0.27\pm
0.04$~\cite{spergel03}.

The main problem in this approach is that it requires numerical
differentiation of noisy experimental data $r_{exp}(z)$. For any set of
experimental data this procedure gives $r'_{exp}(z),\;r''_{exp}(z)$ with
many non-realistic fluctuations which have large amplitude in comparison
with average values. For this reason it is necessary to use smoothing of
initial experimental data. The simplest smoothing procedure is fitting.
However, fitting precision is completely determined by precision of the
fitting function. Several studies have examined different fitting
functions. For example, \cite{turner98} consider the polynomial fit of
$r(z)$ and \cite{weller02} the polynomial fit of
$\tilde\varepsilon_{vac}(z)$. Three parametric fitting function for apparent magnitude
$m(z)$ is suggested in \cite{padmanabhan02}. The authors of \cite{sahni03} proposed
specially constructed fitting function reflected some properties of $r(z)$.
The model dependent approaches were considered in \cite{p99},
\cite{maor02}. The model independent approach where $r(z)$ was fitted
locally with a polynomial fitting is considered in \cite{daly}. In our
opinion, the problem of constructing the fitting function has a unique
solution. In what follows we suggest physically and mathematically motivated function
allowing one to perform the model independent analysis of experimental data
and to constrain the dark energy equation of state.

\section{Motivated fitting function}
We suggest the {\it motivated fitting function} (MFF) which has the
following features:
\begin{itemize}
  \item MFF can be derived from Einstein equations
        {\it i.e.} MFF is not a phenomenological function;
  \item MFF is model independent;
  \item a regular mathematical procedure transforms MFF into
        formally exact solution of Einstein equations;
  \item for any cosmological solution $a(z)$ with $z\le 2$, MFF approximates
        $a(z)$, $a'(z)$, $a''(z)$ with any given precision.
\end{itemize}
With this function, one can increase precision of the model independent approximations
of $r_{exp}(z)$ and two of its derivatives $r'_{exp}(z),\;r''_{exp}(z)$ systematically.
As it is turned out, MFF is completely determined by the following
condition: the only function we need to approximate is scale
factor $a(t)$. It is important that no any other approximation is required.
Let us write cosmological solution and time-redshift relation
\begin{equation}
 \displaystyle
a(x)=1+x+\sum_{n=2}^N a_nx^n,\qquad x=H_0(t-t_0),
\label{a}
\end{equation}
\begin{equation}
\displaystyle x+\sum_{n=2}^N a_nx^n=-\frac{z}{1+z}.
\label{a1}
\end{equation}
The integral (\ref{in}) is approximated as:
\begin{equation}
\displaystyle r_{exp}(z)=-\int_0^{x_z}\frac{dx}{\displaystyle 1+x+\sum_{n=2}^N a_nx^n},
\label{ina}
\end{equation}
where $x_z\equiv x(z)$ is an exact solution of (\ref{a1}) with $x\to 0$ when $z\to
0$. The polynomial coefficients $a_n$ are considered as fitting
parameters. It is obvious that when $N\to \infty$ the polynomial
coefficients can be expressed through derivatives of the cosmological solution $a_n \to
a^{(n)}/n!$ at the moment of observation. In this limit (\ref{a}) --
(\ref{ina}) are formally exact equations. Unfortunately, one
cannot reach this limit in analyzing current experimental data.
For example, the first three coefficients approach the first three
derivatives starting with $N=8$. However, the different cosmological solutions
are approximated with a good precision staring with $N=4$.

One of the great advantages of MFF approach is that the integral can be
expressed through elementary functions after some redefinition of the fitting parameter set
$a_n$: $a_2$, $a_3$, $a_4\ldots\to$, $a$, $b$, $c\ldots$ .
(The redefinition procedure will be described later.)
The only non-trivial point is that the fitting procedure must include
exact solution of equation (\ref{a1}). However, one can use the
fact that the function $x_z\equiv x(z)\equiv x(\tilde z)$ has the
argument $\tilde z=z/(1+z)<1$. So, there exists an iteration
procedure for this function which is always reduced with any given
precision.

In spite of the fact that the result of integration (\ref{ina})
becomes cumbersome with $N\to \infty$, there are no fundamental difficulties
in working with MFF in case of arbitrary $N$. Besides, explicit calculation of (\ref{ina})
is not necessary. The fitting computer program (which is included into
Origin 7 \cite{origin7}) allows us to perform numerical calculation
of (\ref{ina}) during fitting procedure. This approach becomes efficient
starting with $N\geq 5$.

To perform these calculations we need to restrict polynomial order $N$ to
a minimal one. This restriction is fixed by the following
two conditions. The first condition is that the polynomial should
reproduce $a(x)$ and its two derivatives $\dot a, \;\ddot a$ with
respect to $x=H_0(t-t_0)$. The second is that the fitting curve
should be stable against increasing the polynomial order. The
second condition can be satisfied during fitting procedure. To
meet the first condition one needs to study the mathematical
features of the approximation.
We can find the vacuum equation of state from Einstein equations
\begin{equation}
\begin{array}{c}
\displaystyle \tilde \varepsilon_{vac}(x)=
\frac{\dot{a}^2}{a^2}+\frac{\Omega-1}{a^2}-\frac{\Omega_M}{a^3},
\\[7mm]
\displaystyle \tilde
p_{vac}(x)=-\frac13\left(2\frac{\ddot{a}}{a}+\frac{\dot{a}^2}{a^2}+\frac{\Omega-1}{a^2}\right)
\label{eqvac}
\end{array}
\end{equation}
only if we have a good approximation for the first and second derivatives of $a(x)$.
The parametric representation of the equation of state
$\tilde \varepsilon_{vac}=\tilde \varepsilon_{vac}(a),\;\tilde p_{vac}=\tilde p_{vac}(a)$
is derived by  substituting $x=x(a)$ into (\ref{eqvac}).

We studied polynomial approximations of cosmological
solutions for quint\-essence models $\tilde \varepsilon_{vac}=w\tilde
p_{vac}$ with $-1.6\le w\le -0.8,\; \Omega_{DE}=0.73\pm0.04$.
During numerical experiments it has been shown that {\it $N=5$ MFF approximates
any cosmological solution with relative error $|\delta
\varepsilon_{vac}(x)/\varepsilon_{vac}(x)|\le |\delta
p_{vac}(x)/p_{vac}(x)| \le 0.01$ for $z\le 2$.}
If we take $N=4$ (with 3 fitting parameters), we have $|\delta
\varepsilon_{vac}(x)/\varepsilon_{vac}(x)|\sim 0.03,\; |\delta
p_{vac}(x)/p_{vac}(x)|\sim 0.1$.

High precision and universality of the polynomial approximation
of the scale factor and its derivatives means that this
approximation does not represent any model. So, {\it
approximated polynomial with the coefficients obtained by
experimental data fitting can be considered as a model
independent cosmological solution}.

Let us write approximated function $r(z)$ for $N=4$ explicitly.
Polynomial coefficients are redefined in the following way:
\begin{equation}
\begin{array}{c}
\displaystyle a(x)=1+x+a_2x^2+a_3x^3+a_4x^4 \equiv
\\[1mm]
\displaystyle
(ax^2+bx+1)(cx^2+dx+1),
\\[6mm]
\displaystyle a_4=ac,\qquad a_3=ad+bc,
\\[1mm]
\displaystyle a_2=a+bd+c,\qquad d=1-b.
\end{array}
\label{par}
\end{equation}
Integrating (\ref{ina}) gives
\begin{equation}
\begin{array}{c}
\displaystyle r(z)=\frac{1}{2\left[(ad-bc)(d-b)+(a-c)^2\right]}\left[ (ad-bc)
\ln\left|\frac{ax_z^2+bx_z+1}{cx_z^2+dx_z+1}\right|+\right.
\\[6mm]
\displaystyle \left.\frac{2a(a-c)-b(ad-bc)}{\sqrt{|b^2-4a|}}J(x_z;a,b)+
\frac{-2c(a-c)+d(ad-bc)}{\sqrt{|d^2-4c|}}J(x_z;c,d)\right];
\\[6mm]
J(x_z;\alpha,\beta)=\displaystyle \left\{
\begin{array}{ll}
\displaystyle 2\left(\arctan
\frac{\beta}{\sqrt{4\alpha-\beta^2}}-\arctan
\frac{\alpha x_z +\beta}{\sqrt{4\alpha-\beta^2}}\right),\qquad & \beta^2<4\alpha,
\\[6mm]
\displaystyle \ln \left|\frac{(2\alpha x_z+\beta + \sqrt{\beta^2-4\alpha})(\beta -
\sqrt{\beta^2-4\alpha})}{(2\alpha x_z+\beta - \sqrt{\beta^2-4\alpha})(\beta+
\sqrt{\beta^2-4\alpha})}\right|, \qquad & \beta^2>4\alpha.
\end{array}
\right.
\end{array}
\label{ff}
\end{equation}
Equation (\ref{a1}) can be solved through the following iteration procedure
\[
\begin{array}{c}
\displaystyle x_{z(0)}=-\frac{z}{1+z}, \qquad
x_{z(k+1)}=-\frac{z}{1+z}-\sum_{n=2}^4 a_nx_{z(k)}^n,
\\[2mm]
\displaystyle
 0\leq k\leq K
\end{array}
\]
where $K$ is the number of iterations with $|x_{z(K+1)}-x_{z(K)}|<10^{-10}$.

\section{Experimental data fitting}
As we assume, MFF can be useful in analyzing SNAP data. However, in this work we
consider current data to illustrate suggested MFF method. Namely, we consider 78 SNe
data set (Table 3 from \cite{daly}) and 20 RGs
data set (joint fit in Table 1 from\cite{daly}) for $z_{max}=1.79$. Some of
the SNe were observed more than once by independent
observations (the corresponding events will have greater weights).
In this work we do not average out this events, so we consider
the array of 112 points.

\begin{figure}
    \includegraphics[width=10cm]{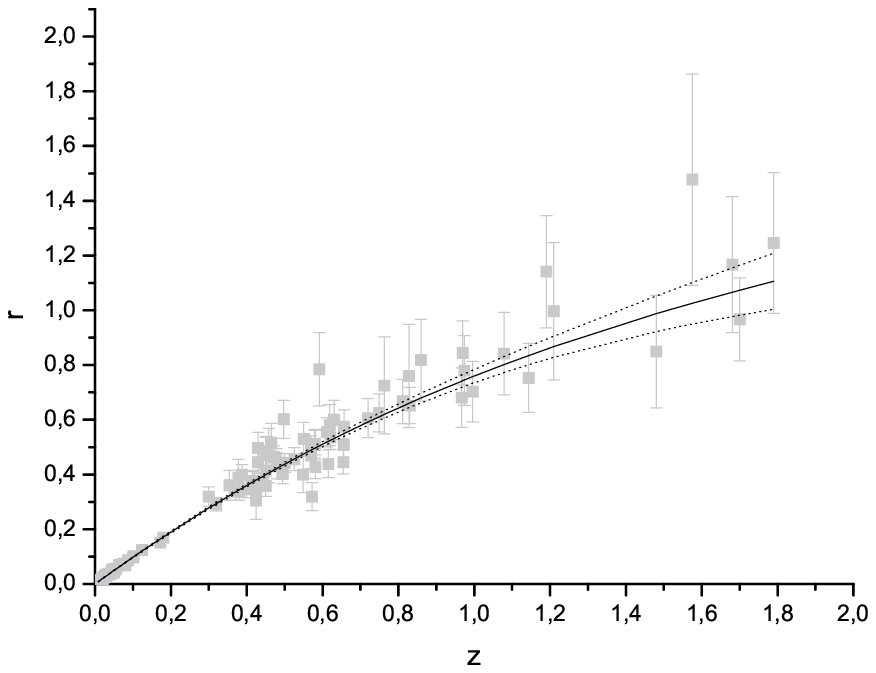}
    \caption{Dimensionless coordinate distance $r(z)$ as a function of $z$. The solid
    line is the best fit. The dotted curves indicate CL68\%. Experimental data are taken
    from \cite{daly} (full data set). }
\label{snapfig1}
\end{figure}

The fitting result is shown in Figure \ref{snapfig1}. We used
instrumental weighting and Levenberg-Marquardt method
\cite{origin7}. The obtained parameters are
\begin{equation}
\begin{array}{c}
\displaystyle a=0.46183^{-0.12089}_{+0.77571}\;,\qquad
b=0.53834^{+0.82220}_{-0.47711}\;,
\\[3mm]
\displaystyle c=-0.47905^{+0.94034}_{-0.87797}\;\qquad
\chi_0^2/\rm{DoF}=1.07499^{+0.02390}_{+0.03147}\; .
\end{array}
\label{fit1}
\end{equation}
Upper and lower values in (\ref{fit1}) correspond to parameters for upper and lower
curves which bound CL68\% region.

This data set has been fitted by MFF with $N=3$ and $N=5$. It is
turned out that fitting curves are almost coincide in the region
where $z<1$. In the region $z>1$ MFF with $N=3$ does not work satisfactorily
(for example the corresponding $\chi^2$ is greater than those in (\ref{fit1}).)
On the contrary in the region $z>1.5$ MFF with $N=5$ is very
sensitive to single experimental points and has non-realistic wave-like behaviour.
For this reason MFF with $N=4$ (3-parametric function) is optimal for
the analysis of the experimental data. Note, that for the SNAP data, optimal $N$
will be greater (most likely $N=5$).

One can see from (\ref{fit1}) that we have a set of fitting
curves with almost the same $\chi^2$.
Every fitting curve corresponds to a cosmological solution and we need to
find the true solution. To do this we attempted to average
experimental data. We considered two types of averaging. In the first type
we calculated average values summing data on the different small $\Delta
z$ regions. In the second type we divided $z$ axis into three or four
regions, fitted data independently on each region and then
combined resulting curves using MFF. These methods gave almost the same
result: averaged curves lie inside CL68\% not far from initial
fitting curve. After the long numerical experimentations we concluded
that initial fitting curve is a better candidate for reflecting
the true cosmological solution.

\section{Vacuum equation of state}
From the fact that geometrical set of averaged data points lie not far from initial
fitting curve it follows that MFF with $N\geq 4$ can
reconstruct structural features of the true
solution at the different redshift regions. This allows us to
obtain the first derivatives $r'(z), r''(z)$ directly from analytical
expression for MFF.  For this reason $r_{fit}(z)$ can be regarded
as the mean values and $\Delta r_{syst}(z)=|r_{exp}(z)
- r_{fit}(z)|$ as the systematic errors. So, each experimental
point can be parameterized by the statistical confidence parameter
\[
\displaystyle w(z)=\left|\frac{\Delta r_{stat}(z)+\Delta
r_{syst}(z)}{r_{fit}(z)}\right|,
\]
where $\Delta r_{stat}(z)$ is the statistical error (denoted in \cite{daly} as
$\sigma(y_i)$). We suggest to use this parameter for estimating
various physical quantities $\mathcal{F}_{fit}(z)$ calculated
using fitting curve. Here we can introduce conventional errors $\Delta
\mathcal{F}(z)=w(z)|\mathcal{F}_{fit}(z)|$.

\begin{figure}
    \includegraphics[width=10cm]{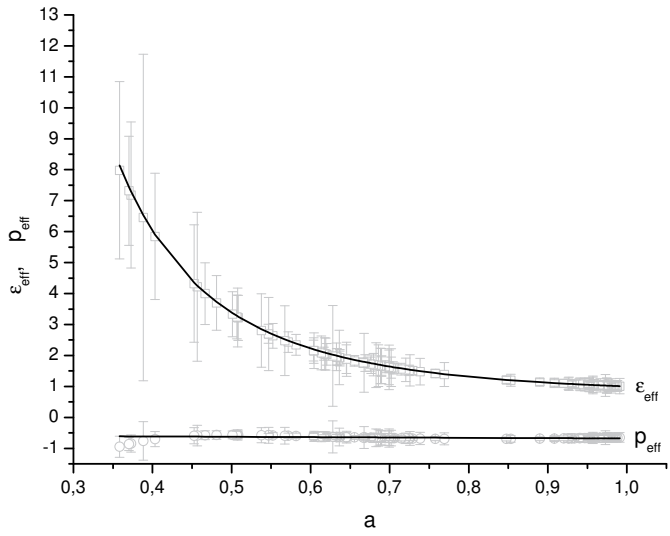}
    \caption{Effective energy density and pressure as a function of $a$ obtained
    through model independent analysis of experimental data using MFF.
    Fitting curves correspond to the model (\ref{mod}) with parameters from  (\ref{fitmod}).}
\label{snapfig2}
\end{figure}

Let us turn to vacuum equation of state and consider the function
\begin{equation}
\begin{array}{c}
\displaystyle \tilde \varepsilon_{eff}(a)\equiv \tilde
\varepsilon_{vac}(a)+\frac{\Omega_M}{a^3}-\frac{\Omega-1}{a^2}=\frac{1}{(r'_{fit})^2},
\\[6mm]
\displaystyle \tilde p_{eff}(a)\equiv \tilde
p_{vac}(a)+\frac{\Omega-1}{3a^2}=-\frac{1}{(r'_{fit})^2}-\frac{2r''_{fit}}{3a(r'_{fit})^3},
\end{array}
\label{epeff}
\end{equation}
which is derived using data fit (see Figure \ref{snapfig2}).
When $z\sim2$ it is necessary to use both functions in (\ref{epeff}).
It should be emphasized that {\it functions plotted in Figure \ref{snapfig2}
were obtained in a model independent way}.

Since $\tilde p_{eff}(a)$ changes slowly it is reasonable to
suppose that flat Universe model with a cosmological constant
could be a possible cosmological model. More detailed information can be
obtained through fitting $\tilde \varepsilon_{eff}(a), \;\tilde p_{eff}(a)$
with the following model
\begin{equation}
\begin{array}{c}
\displaystyle \tilde \varepsilon_{eff}(a)=
\frac{\Omega_{DE}}{a^n}+\frac{\Omega-\Omega_{DE}}{a^3}-\frac{\Omega-1}{a^2},
\\[5mm]
\displaystyle \tilde p_{eff}(a)=\frac{(n-3)\Omega_{DE}}{3a^n}
+\frac{\Omega-1}{3a^2}
\end{array}
\label{mod}
\end{equation}
which has three fitting parameters $\Omega_{DE}, \;n,\; \Omega$.
As a result we have
\begin{equation}
\begin{array}{c}
\displaystyle \Omega_{DE}=0.64290\pm 0.07385,\qquad n=-0.14445\pm
0.23847,
\\[4mm]
\displaystyle \Omega=0.98801\pm0.10433, \\[4mm]
\displaystyle w\equiv \frac{\tilde p_{vac}(a)}{\tilde
\varepsilon_{vac}(a)}=-1.04815\pm 0.07949,
\\[4mm]
\displaystyle \chi^2=0.04584.
\end{array}
\label{fitmod}
\end{equation}
For non-flat LCDM model ($n\equiv 0,\; \Omega_{DE}\equiv \Omega_\Lambda$) we have:
\begin{equation}
\begin{array}{c}
\displaystyle \Omega_{\Lambda}=0.68973\pm 0.01897,\qquad  \Omega=1.05293\pm 0.03068,
\\[2mm]
\displaystyle \chi^2=0.04753.
\label{fitmod1}
\end{array}
\end{equation}
One can see from (\ref{fitmod}), (\ref{fitmod1}) that these
results are in a good agreement with WMAP experimental data
\cite{spergel03}. We hope that presented MFF method can provide
an independent check to the traditional cosmological tests.

\newpage

\begin{thebibliography}{99}
\bibitem{r98} A.~G.~Riess, A.~V.~Filippenko, P.~Challis {\it et al.},
Astron.\ J.\ \textbf{116}, 1009 (1998) preprint astro-ph/9805201.
\bibitem{p99} S.~Perlmutter, G.~Aldering, G.~Goldhaber {\it et al.},
Astrophys.\ J.\ \textbf{517}, 565 (1999) preprint astro-ph/9812133.
\bibitem{d02} R.~A.~Daly, E.~J.~Guerra, Astron.\ J.\, \textbf{124}, 1831,
(2002).
\bibitem{peebles02} P.~J.~E.~Peebles, B.~Ratra, preprint astro-ph/0207347,
(2002).
\bibitem{turner02} M.~S.~Turner, Int.\ J.\ Mod.\ Phys.\ A \textbf{17S1},
180 (2002) preprint astro-ph/0202008.
\bibitem{aldering02} G.~Aldering, C.~Akerlof, R.~Amanullah {\it et al.} [SNAP
Collaboration], preprint astro-ph/0209550.
\bibitem{daly} R.~A.~Daly and S.~G.~Djorgovski, preprint astro-ph/0305197.
\bibitem{spergel03} D.~N.~Spergel, L.~Verde, H.~V.~Peiris {\it et al.},
preprint astro-ph/0302209.
\bibitem{turner98} D.~Huterer, M.~S.~Turner, preprint astro-ph/0103175.
\bibitem{weller02} J.~Weller, A.~Albrecht, Phys.\ Rev.\ D \textbf{65},
103512 (2002) preprint astro-ph/0106079.
\bibitem{maor02} I.~Maor, R.~Brustein, J.~McMahon {\it et al.}, Phys.\
Rev.\ D \textbf{65}, 123003 (2002) preprint astro-ph/0112526.
\bibitem{sahni03} T.~Saini {\it et al. } Phys.\ Rev.\ Lett.\ \textbf{85} 
(2000) 1162, preprint astro-ph/9910231.
\bibitem{padmanabhan02} T.~Padmanabhan, T.~Roy Choudhury, preprint astro-ph/0212573.
\bibitem{origin7} Origin 7, OriginLab 2002.
\end{thebibliography}
\end{document}